\documentclass{kluwer} 
\usepackage{graphicx}
\newdisplay{guess}{Conjecture}

\begin{document}                                                                              
\begin{article}
\begin{opening}

\title{Formation and Evolution of E and S0 Galaxies from HST and Keck Studies of 
$z\sim $ 0.3--1 Clusters.}

\author{Garth \surname{Illingworth}}  
\runningauthor{Garth Illingworth}
\runningtitle{Intermediate Redshift Cluster Galaxies}
\institute{UCO/Lick Observatory, Astronomy and Astrophysics Department, 
University of California, Santa Cruz, CA \ \ 95064}
\author{Daniel Kelson}
\institute{Department of Terrestrial Magnetism, 5241 Broad Branch Road, NW,
Washington DC 20015}
\author{Pieter van Dokkum and Marijn Franx}
\institute{Leiden Observatory, P.O. Box 9513, 2300 RA Leiden, The Netherlands}

\date{October 15, 1999}

\begin{abstract}

We have witnessed a dramatic increase over the last five years in results on distant 
galaxies, in large part because of the high resolution imaging capability of HST,
and the multiobject spectroscopic capability of the Keck telescopes. Our program to obtain 
wide-field, multi-color WFPC2 mosaics with HST of intermediate redshift
clusters, and spectroscopic membership and high S/N spectroscopy with LRIS on Keck, 
has provided new insights into the nature of elliptical and S0 galaxies in the cluster
environment over a wide range of densities.  In particular, most ellipticals, and 
a significant fraction of the S0 population, have large luminosity-weighted ages,
suggesting that their stellar populations were formed at redshifts beyond $z\sim 2$,
though the existence of substantial numbers of major mergers in MS 1054-03 at $z=0.83$ 
suggests that final assembly of such galaxies may not have occurred until much later for a
significant fraction of early-type galaxies.

\end{abstract}


\end{opening}

\section{Introduction}  

HST WFPC2 imaging and Keck LRIS spectroscopy have helped to revolutionize
our understanding of high redshift galaxies. In particular, they have 
played a central role in the study of the evolution of early-type 
galaxies in intermediate redshift clusters. 

\begin{figure}
\centerline{(See 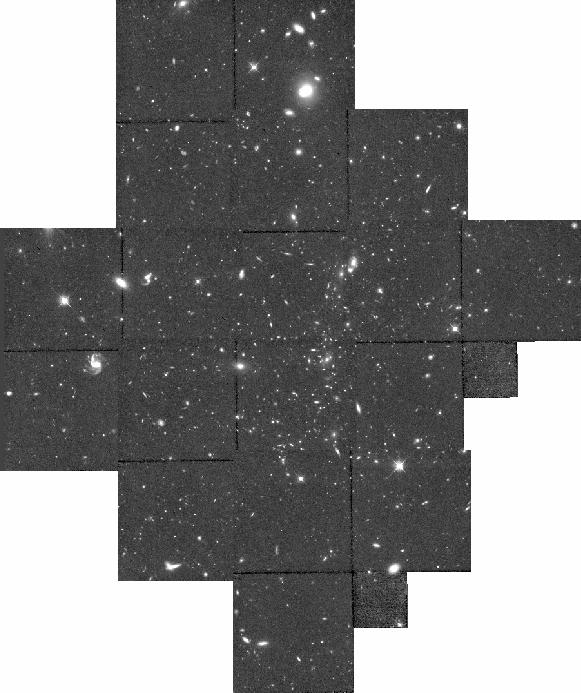)}
\caption{The HST WFPC2 36-orbit, six-pointing two-color mosaic of the $z=0.83$
X-ray cluster MS1054--03.  This image covers $5'\times7'$, i.e., 
out to $\sim 2h_{65}^{-1}$ Mpc radius.}
\end{figure} 

Over the last four years we have been carrying out HST WFPC2
imaging and Keck multislit spectroscopy of three massive, X-ray selected clusters
ranging in redshift from CL 1358+62 at $z=0.33$,  to MS 2053-04 at $z=0.58$, and  to MS 1054-03
at$z=0.83$.  A key factor that distinguishes
this work from other programs with HST is that we have used multiple pointings
to cover a wide area around the cluster.  
The advantage of this approach for many issues relating to cluster and early-type 
galaxy evolution became clear as the project developed.  The results from this program
have resulted in two dissertations (Daniel Kelson: UCSC 1998, and Pieter van 
Dokkum: Groningen 1999).   

We have used multiple pointings on HST to image regions with WFPC2  
that are typically 5--7$'$ $\times$ 5--7$'$ in size.   This corresponds to 
a field size of roughly 1.5$h_{65}^{-1}$ -- 2$h_{65}^{-1}$ Mpc, centered on the clusters.   
An example of the mosaic HST images for one of our cluster fields is shown in Figure 1.
The fields were imaged with two filters, F606W and F814W. Spectroscopy was carried out using 
the Keck LRIS multi-object spectrograph, with a field comparable to that of the HST mosaics.

The scientific goals of this program have included  constraining the evolution of
early-type galaxies, identifying the galaxy characteristics across the cluster 
as a function of environment and density, analysis of the cluster's spatial and velocity
structure,  and determining the mass distribution through weak lensing.
The primary tools that we have used for characterizing the evolution of the 
early-type galaxy population have been the fundamental plane, the color-magnitude 
relations, and more recently, absorption line strengths.  

\begin{figure}
\centerline{
\includegraphics[width=10pc]{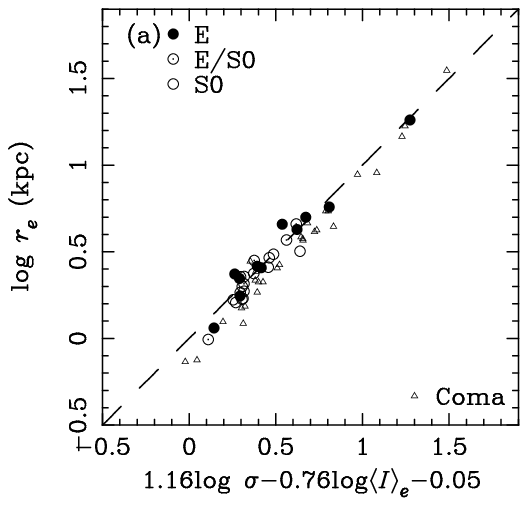}\qquad
\includegraphics[width=10pc]{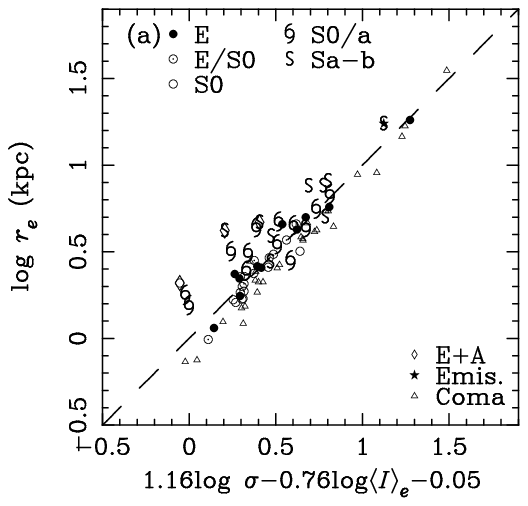}
}
\caption{a)The fundamental plane of 30 E/S0 galaxies in Cl 1358+62 at
$z=0.33$, showing the very tight relation (comparable to that in Coma).  The
intrinsic scatter is only 14\% in $r_e$.   The (small) offset to Coma is evolutionary
brightening, since the $(1+z)^4$ surface brightness dimming has been removed.
The full sample of 53 galaxies, including the early-type spirals and E+A galaxies,  
is shown in (b). The offsets for these galaxies are as expected for a 
younger luminosity-weighted age for their stellar populations (see Kelson et al. 2000).}
\end{figure}

\section{Fundamental Plane}

The fundamental plane is a tight relation in early-type galaxies between velocity dispersion,
effective radius and mean surface brightness that is described by 
$r_e \propto \langle I_e \rangle^{-0.83}\sigma^{1.20}$.  With reasonable assumptions about homology,
this implies that $M/L \propto M^{0.25}$. The very small scatter, $\pm$23\% in Coma 
in $M/L_V$, makes it an ideal tool for establishing the evolution of $M/L$ with redshift.  
Since age is related to $M/L$ for early-type galaxy populations, 
constraints can be put on their (luminosity-weighted) ages and age distributions.
This requires high S/N data, and attention to minimizing systematic 
errors.  For $z\sim $ 0.3--1 clusters, HST images are needed to derive $r_e$ and $I_e$, 
while  8-10 m telescopes are needed for $\sigma$ (6-8 hour integrations were used at Keck at $z=0.83$).

\begin{figure}
\centerline{\includegraphics[width=\linewidth]{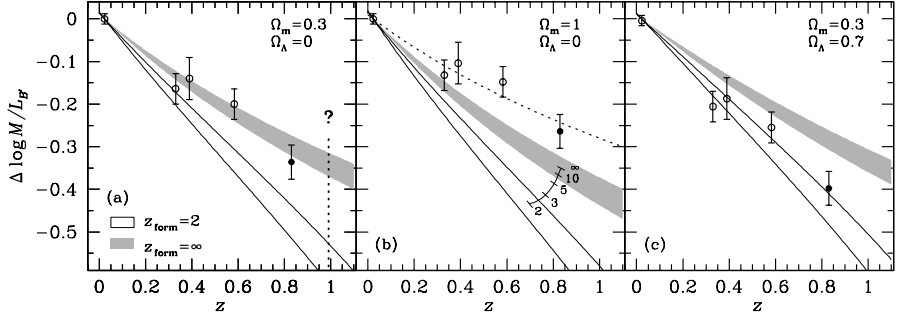}}
\caption{Evolution of $M/L_B$ with redshift from the fundamental plane measurements
in Coma at $z=0.02$, CL 1358+62 at $z=0.33$, CL 0024+16 at $z=0.39$, MS2053-04 at $z=0.58$, 
and MS 1054-03 at $z=0.83$, for an (a) open $\Omega_m = 0.3$, a (b) flat, matter-dominated 
$\Omega_m =1$, and (c) a lambda cosmology $\Omega_{\Lambda}=0.7, \Omega_m =0.3$.
Single-burst model predictions are shown for a Salpeter IMF  ($x=1.35$) and a
range of metallicities, as indicated by the spread in $\Delta {\rm log} M/L_{B'}$
for a given $z_{form}$.  The sensitivity to the IMF is shown in (b) 
as a dotted line ($z_{form} = \infty$ and a steep IMF with $x=2.35$), as is the 
effect of reducing $z_{form}$ from $z_{form} = \infty$.   The data favor high 
formation redshifts and a low $q_0$, but this result is sensitive to the IMF.}
\end{figure}

For CL 1358+62, high S/N spectra were obtained of 53 galaxies 
in the HST mosaic, resulting in the derivation of the fundamental plane shown in Figure 2
(Kelson et al. 2000).
These data made it possible to derive a fundamental plane at $z=0.33$ that is comparable to
Coma.  The scatter in $M/L$ in the E/S0 population in CL 1358+62 is just 16\%.
The offset from Coma constrains the luminosity-weighted ages to $z>1$, and the 
consistency of the scatter in the fundamental plane and the color-magnitude relation 
constrains the luminosity-weighted age dispersion to be $<$15\%.

Based on our preliminary fundamental plane measurements at higher redshift  
(Kelson et al. 1997, and  van Dokkum et al. 1998a), we have used the offsets in $\Delta M/L$ 
from Coma  to set even stronger constraints, on the luminosity-weighted ages, and on the 
cosmology, as shown in Figure 3 (van Dokkum et al. 1998a).
An Einstein-de Sitter ($\Omega_m =1$) is only consistent with the data if the 
IMF is substantially steeper than Salpeter.    These data give 
$\Delta$log $M/L_B \propto -0.4z$.  Tighter constraints would be 
placed by observing even slightly higher redshift clusters, though the spectroscopy 
becomes very time-consuming, even with  Keck!

\section{Color-Magnitude Relations}

The color-magnitude (CM) relations provide a complementary approach to 
the fundamental plane for characterizing the evolution of early-type
galaxies.  They provide an independent means of constraining $\Delta t/t$.  
The use of such relations brings a 
benefit in that larger samples can be readily obtained (redshifts are easier 
than velocity dispersions), 
but the use of these relations for constraining the star formation  
history of old stellar populations does require very accurate photometry.
This can be done with HST data, which also enables the required morphological 
determinations to be made.   The value of such
data can be seen in Figure 4 (See 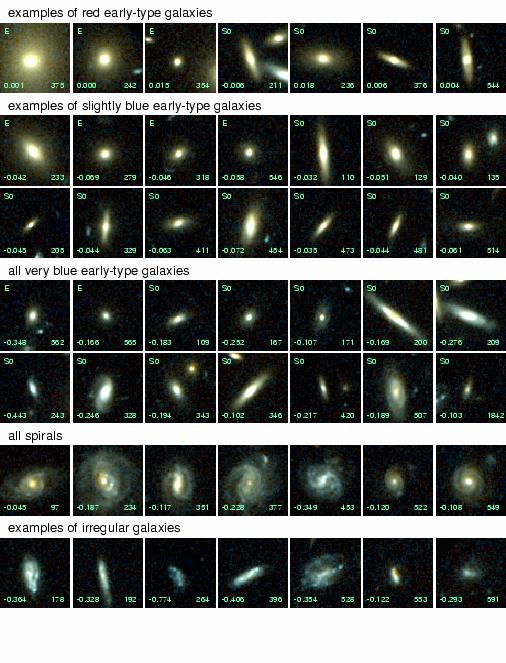) where color images from HST show the range
of morphological and structural information that can be obtained in CL1358+62 at $z=0.33$.

An extensive CM analysis of the large sample of members with spectroscopic 
redshifts in CL 1358+62 was carried out by van Dokkum et al. (1998b). 
The key to this analysis was the very accurate photometry that could be carried 
out on the WFPC2 images.  For galaxies with $V_z < 21$, the photometric uncertainty was 
demonstrated to be 0.009 mag, rising to only 0.017 mag for fainter objects.
With 194 members over a radius of 4.6 arcmin, or 1.2 $h_{65}^{-1}$ Mpc,
the colors of galaxies could be investigated as a function of radius. As shown in Figure
5, the elliptical galaxy population shows very small scatter at both small and large radii, and
no significant change in scatter with radius (a biweight estimator was used to minimize the effects 
of outliers).
The S0 galaxies likewise show little scatter at small radii, and  are essentially identical to the 
ellipticals, but at larger radii the S0 population is bluer and shows substantially larger scatter:
a factor 2 higher than that of the S0s in the center (or the ellipticals at all radii).
The difference in the mean color (to bluer colors) in the S0 populations is significant at the 
95\% confidence level.  For the ellipticals at all radii, and the S0s in the central regions,
the scatter implies a luminosity-weighted age distribution with $\Delta t/t \sim 0.18$, while the 
luminosity-weighted age distribution for the S0s at $r>118''$ is $\Delta t/t \sim 0.35$, 
with (some) outer S0s experiencing star formation almost to the epoch of observation at $z=0.33$.  

This powerful technique is also being applied to our other clusters.

\addtocounter{figure}{1}
\begin{figure}
\centerline{\includegraphics[width=\linewidth]{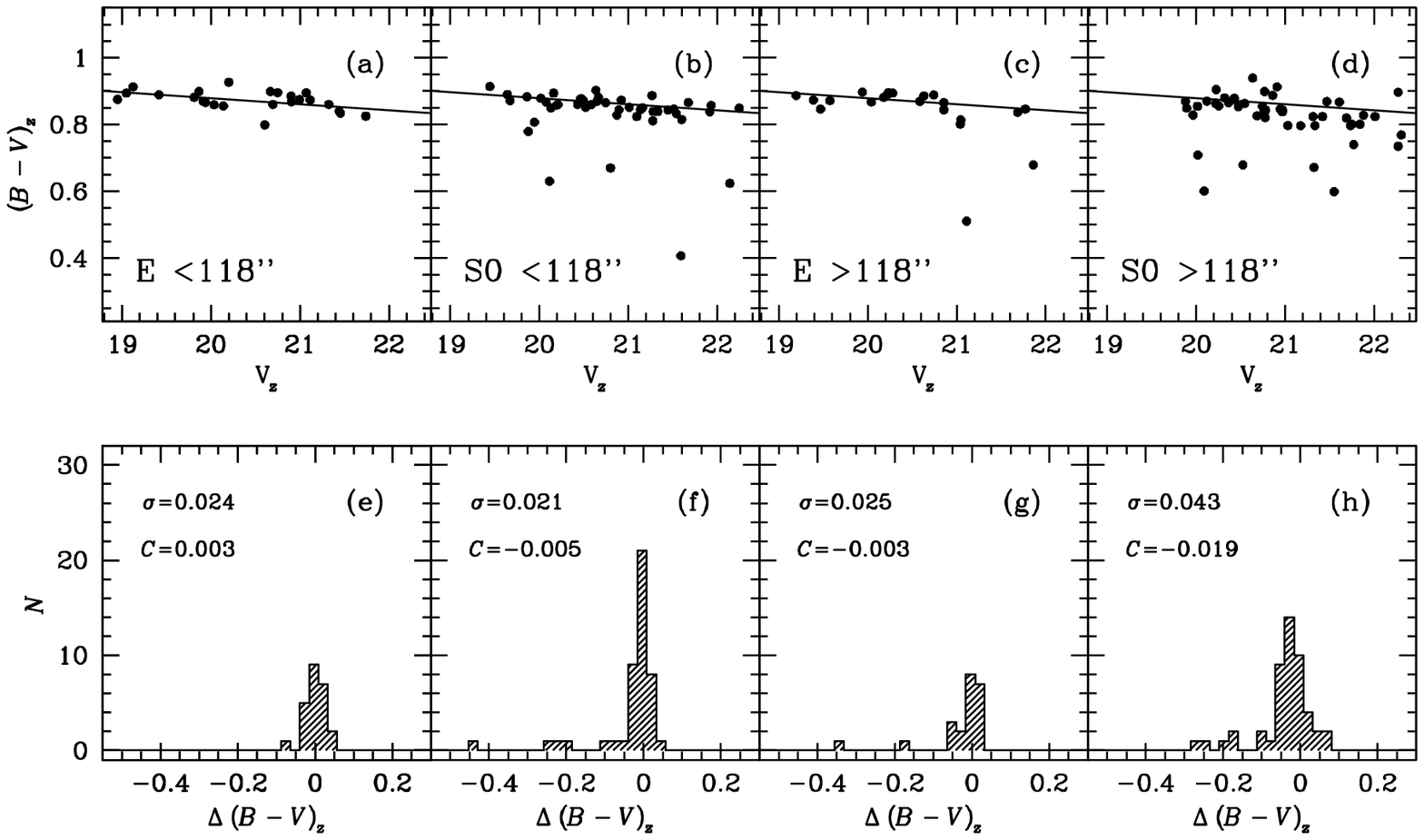}}
\caption{Comparison of the color-magnitude relations
for ellipticals and S0s in CL 1358+62 at $z=0.33$, as a function of radius 
in the cluster.  The scatter in the CM relation of the inner region S0s is
the same as the ellipticals (which show no dependence with radius).
Outside $r=118''$, the scatter in the S0 colors is larger
than that of the ellipticals and inner S0s by a factor 2, and they are bluer in the mean. 
The radius 118$''$ was chosen so as to split the sample into two equal groups. }
\end{figure}

\section{Major Mergers in MS 1054--03}

The value of the large field coverage of our HST imaging was demonstrated again
with our mosaic on MS 1054-03 at $z=0.83$ (Figure 1), when the morphological types of 
the 89 members that resulted from the Keck LRIS spectroscopic magnitude-limited sample
($I< 22.2$) were analyzed.  The morphological types were split 22\% E, 22\% S0, 39\% spiral
and 17\% {\it merger/peculiar}.  To find such a high fraction of luminous
merger/peculiar galaxies in such a rich cluster at such a late time was a surprise.
The morphological types of the most luminous galaxies in MS 1054-03, plus the lower 
luminosity merger/peculiar objects, are shown in Figure 6  (See 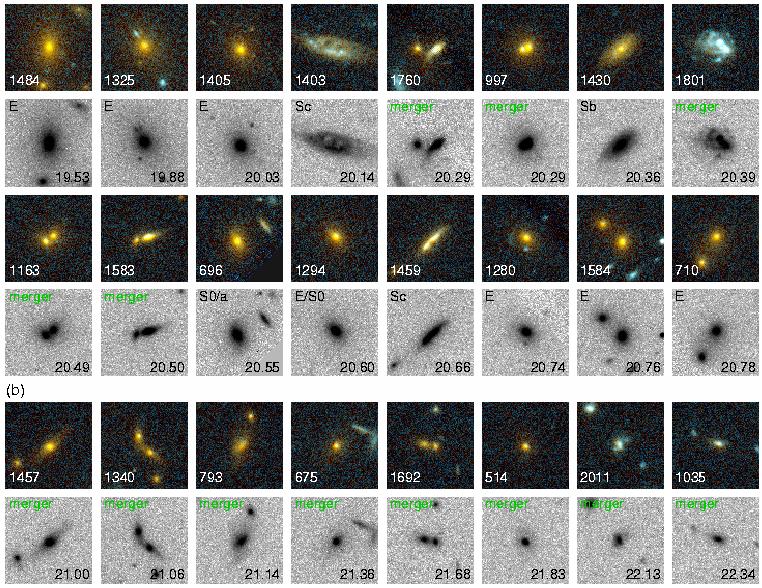).

These results are discussed in more detail in van Dokkum et al. (1999).
The color magnitude diagram shown in Figure 7b demonstrates one of the remarkable aspects  
of the mergers in this cluster -- they are quite red, with no detected 
[O{\sc ii}]\,3727\,\AA{} emission, suggesting minimal ongoing star formation.   
The mergers are also found in the outer parts of the cluster, suggesting that they 
are in infalling clumps (Figure 7a).   
Most of these mergers are likely to evolve into luminous 
($\sim 2L^*$) early-type galaxies, presumably ellipticals, but also possibly S0s.
If confirmed to be generally the case in other rich $z\sim $ 0.5--1  clusters, this
result suggests that the merging rate in the cluster environment is 
changing rapidly with redshift
(possibly as fast as $(1+z)^6$).  This important result is consistent with the predictions
of hierarchical clustering models (see, e.g., Kauffmann 1996).

\addtocounter{figure}{1}
\begin{figure}
\centerline{
\includegraphics[width=10pc]{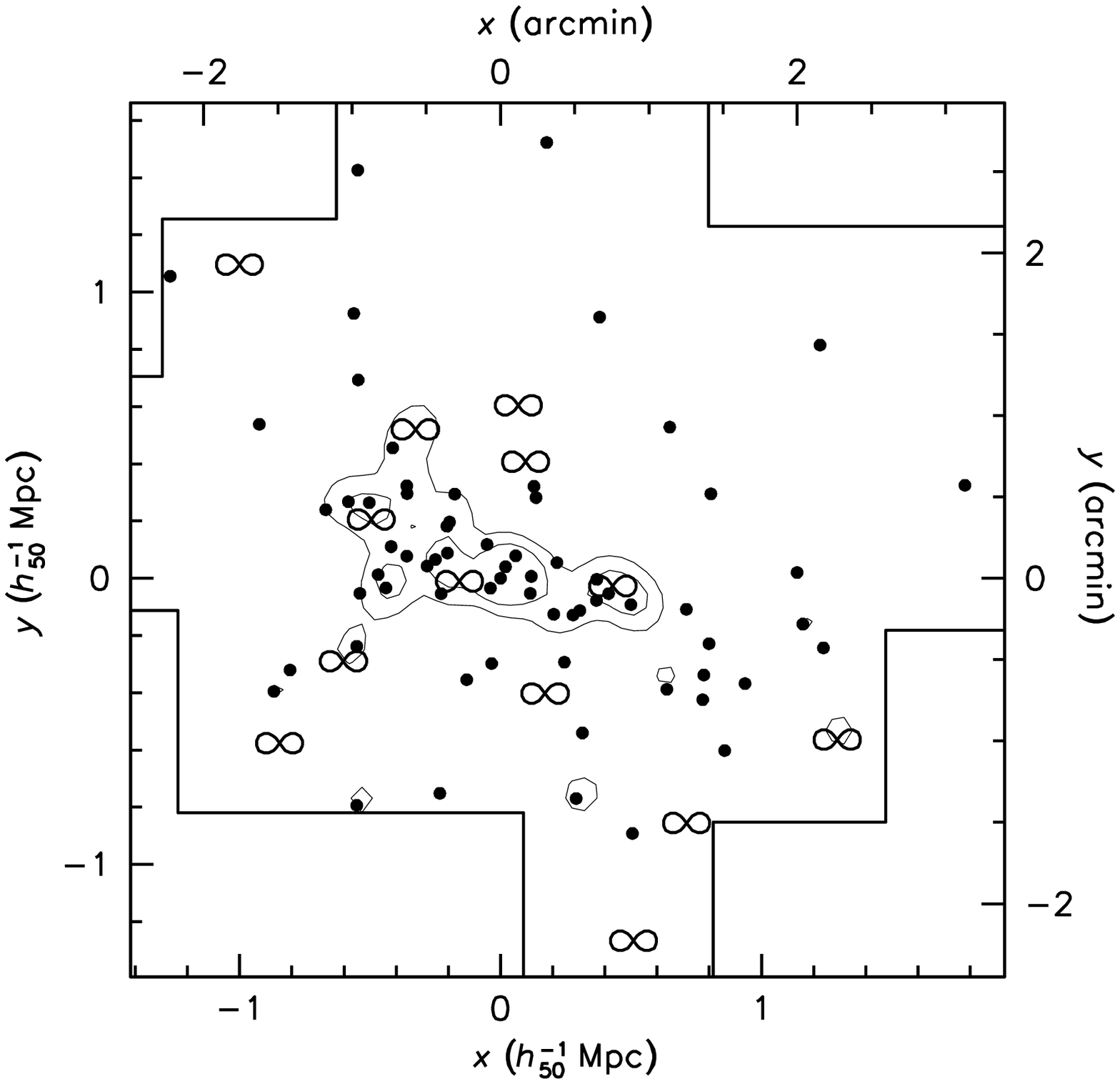}\qquad
\includegraphics[width=10pc]{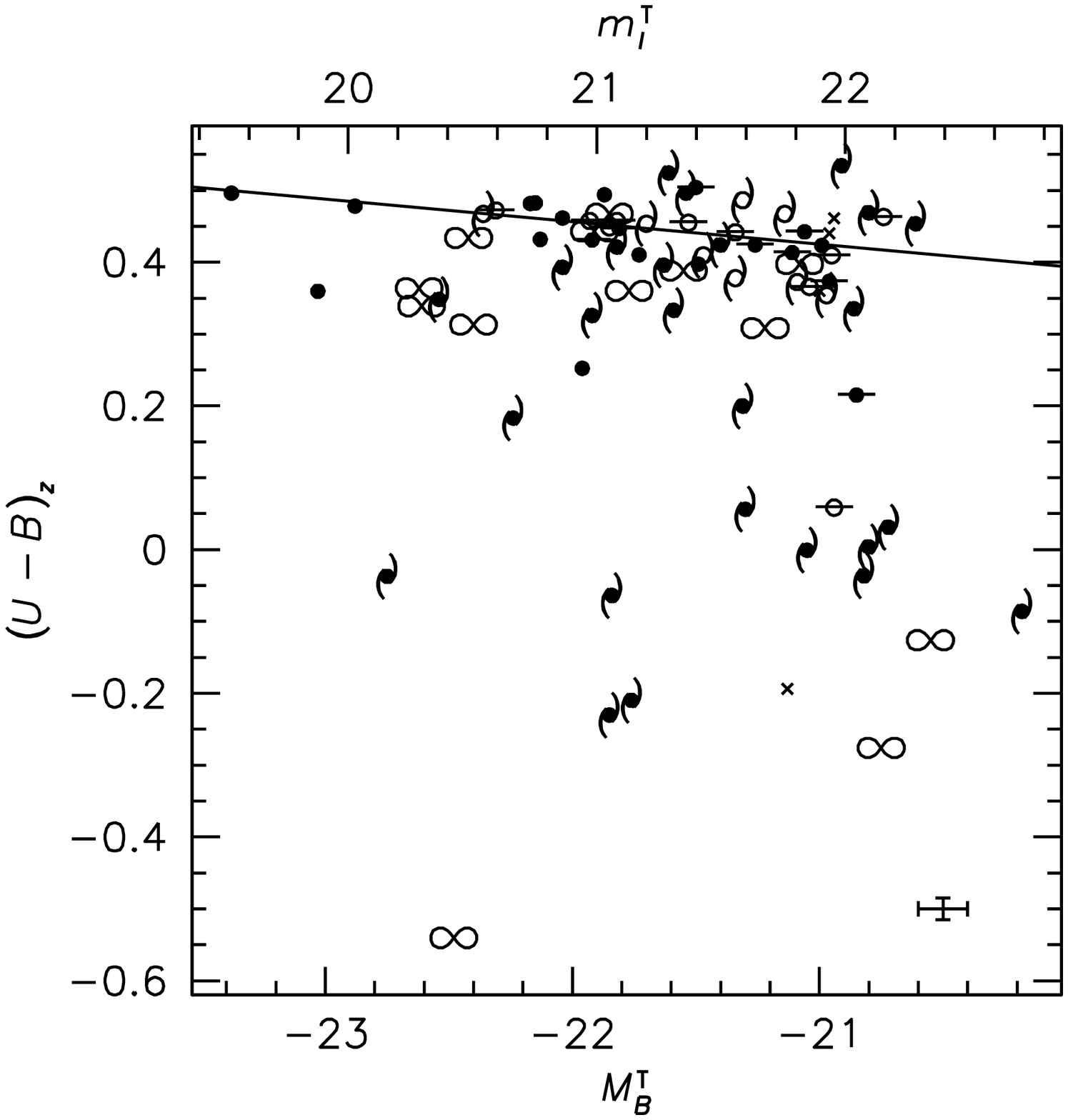}
}
\caption{(a) Spatial distribution of the confirmed cluster members in MS 1054-03, with contours
of the the red galaxy  distribution, and the mergers indicated as $\infty$ symbols.
(b) Color-magnitude relation of the spectroscopically-confirmed 
members of MS1054--3.  Excluding the three very blue mergers, the bulk of the mergers
are only $\sim 0.07$ mag bluer than the CM relation defined by the E/SO
early-type galaxies (solid line).}
\end{figure}

\section{Line Strengths}

We have recently extended our program to encompass line strength measurements.
The high S/N data that were obtained for deriving velocity dispersions
for the fundamental planes in the three clusters are also ideal for deriving
line strengths.  The first results from this program are given in 
Kelson et al. (1999).  An important demonstration of the potential
of the use of line strengths for substantial samples of intermediate-redshift, 
early-type galaxies  can be seen in Figure 8.  The tight correlation between the metal-sensitive 
index C$_2$4668 \AA \ and the restframe $B-V$ color in CL 1358+62 at $z=0.33$ has a slope 
that is consistent with that purely for metallicity variation at constant age, with no need for 
age variation. 
Line strengths will prove to be a significant component of our ``toolbox'' for 
studying distant galaxies.

\begin{figure}
\centerline{\includegraphics[width=12pc]{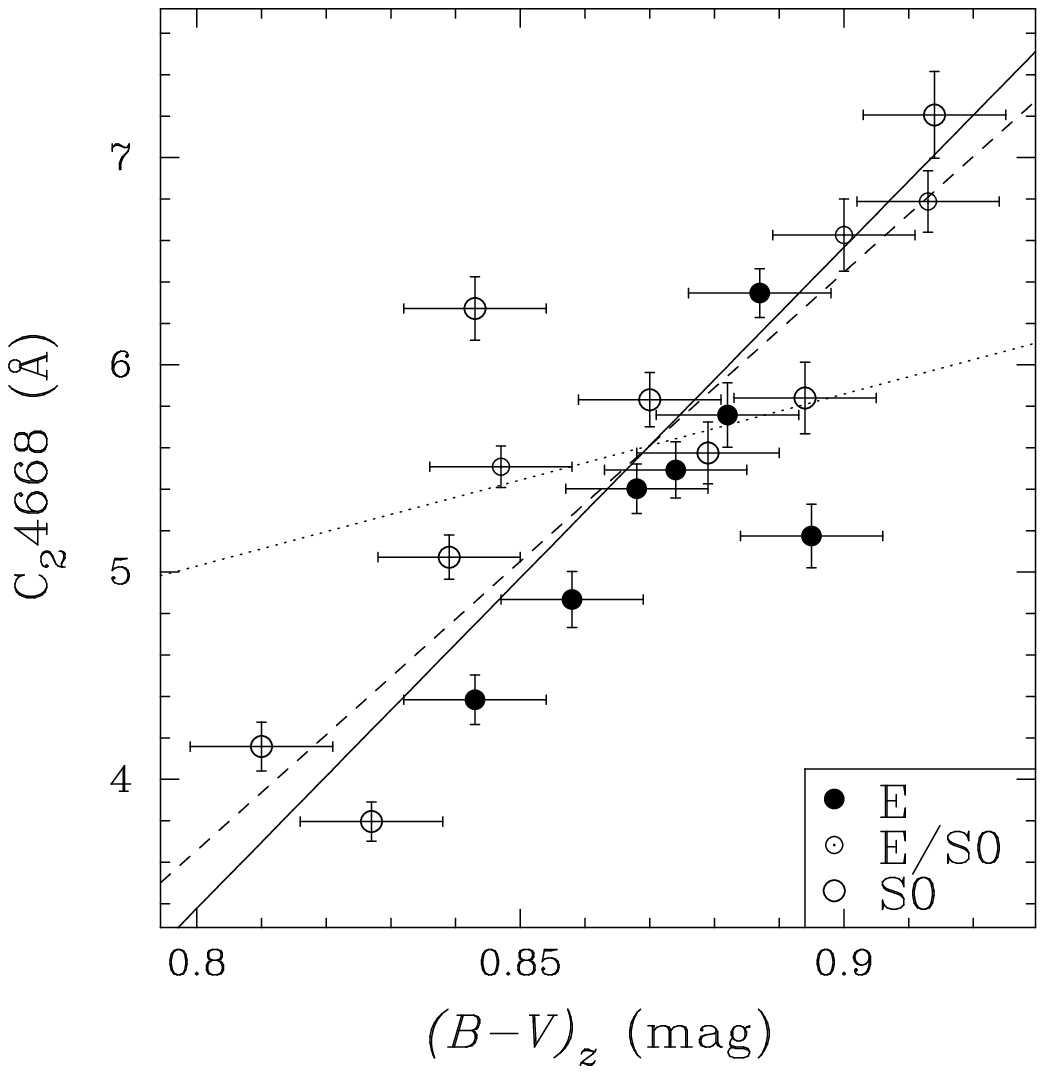}}
\caption{The C$_2$4668 \AA \ index plotted against restframe $B-V$ color for
the E/S0 galaxies in CL 1358+62, with the best fit line (solid). Predictions
from single-burst stellar population models are shown
for (i) age variations at constant Fe/H (dotted), and (ii) metallicity
variations at constant age (dashed).   The data are consistent with metallicity
variations at a fixed age -- see Kelson et al. 1999 for more details.}
\end{figure}
 
\section{Summary}

The fundamental plane and color-magnitude relations (and the forthcoming 
line strength studies) provide complementary approaches to establishing the evolutionary 
history of early-type galaxies.  Already our results show that the 
luminosity-weighted ages of E/S0 galaxies in intermediate redshift ($z\sim $ 0.3--1) 
massive X-ray clusters are old, i.e., $z_f>2$, with small $\Delta t / t \sim $ 0.15--0.18.  
Furthermore, our recent results on the major mergers in MS 1054-03 
indicate that a substantial fraction of the 
E/S0 population could have undergone a final ``assembly'' phase at $z<1$.

\acknowledgements

The lead author is very grateful to David Block for organizing such 
an excellent conference,  and to Margi Crookes for her dedication to 
making everything work well,  and to all those who helped them.  
I would particularly like to thank the Anglo-American Chairman's Fund and 
SASOL for their generous financial support.
Support from STScI grants GO07372.01-96A, GO06745.01-95A, and  
GO05989.01-94A is gratefully acknowledged.

\end{article}
\end{document}